\pdfoutput=1

\documentclass[10pt,conference]{IEEEtran}
\IEEEoverridecommandlockouts

\usepackage{cite}
\usepackage{amsmath,amssymb}
\usepackage{booktabs}
\usepackage{graphicx}
\usepackage{array}
\usepackage{xcolor}
\usepackage[hyphens]{url}
\usepackage[hidelinks]{hyperref}
\graphicspath{{./figures/}}

\begin{document}

\title{Hybrid Classical-Quantum Optimization for CVRPTW: A Systems
Framework with Provable Feasibility and Empirical Hardware Limits}

\author{ \IEEEauthorblockN{Syamasundar Gopasana, Swatantra Singh, Rohan Joydhar, and Shubhashis Sengupta} \IEEEauthorblockA{ \textit{Accenture Labs}\\ India\\ \{syamasundar.gopasana, swatantra.a.singh, rohan.joydhar, shubhashis.sengupta\}@accenture.com } }

\maketitle

\begin{abstract}
Solving large-scale Capacitated Vehicle Routing Problems with Time
Windows (CVRPTW) remains a critical yet computationally intractable
challenge in real-world logistics systems, where strict operational
constraints and scale limit the effectiveness of both exact and
heuristic methods.  At the same time, emerging quantum and
quantum-inspired approaches promise new optimization capabilities, but
their applicability to industrial-scale routing remains unclear.

This paper presents a hybrid optimization architecture that combines
classical Mixed Integer Linear Programming (MILP) for global
feasibility using Gurobi, with vehicle-level Quadratic Unconstrained Binary
Optimization (QUBO) subproblems solved using quantum-inspired Simulated
Bifurcation Machines (Toshiba SQBM+) and the Quantum Approximate
Optimization Algorithm (QAOA) on Amazon Braket backends (Rigetti
Ankaa-3 and IQM Emerald).  The design is motivated by a
systematic analysis showing that monolithic CVRPTW QUBO formulations
are intractable even for small instances due to variable explosion and
unstable constraint enforcement.

The framework is evaluated on five realistic routing benchmarks derived
from real London traffic and location data (10--70 customers) with a
strict validation pipeline enforcing feasibility and monotone
improvement over classical baselines.  SQBM+ achieves verified
fleet-distance reductions of $\mathbf{2.80\%}$ ($7{,}927$~m) for the
25-customer instance and $\mathbf{0.22\%}$ ($1{,}186$~m) for the
55-customer instance.  QAOA experiments across multiple hardware
platforms demonstrate reliable constraint satisfaction only for
subproblems up to 4 customers ($\le 16$~qubits) on current hardware.

Beyond performance, the study provides diagnostic insights showing that
assignment-level time-window tightness, not local route
optimization, is the dominant scalability bottleneck.  These results
position hybrid classical--quantum architectures as a practical
near-term pathway for logistics optimization, while establishing
realistic performance boundaries for emerging quantum systems.
\end{abstract}


\section{Introduction}

The Capacitated Vehicle Routing Problem with Time Windows (CVRPTW)
requires routing a fleet of vehicles to serve geographically distributed
customers under simultaneous capacity and per-customer service-time-window
constraints.  It is NP-hard in the general
case~\cite{toth2014vehicle}.  As the number of customers and vehicles
grows, the number of feasible route-assignment combinations multiplies
rapidly: each additional delivery point expands both the routing search
space and the constraint enforcement burden, causing solver complexity
to escalate super-linearly (Fig.~\ref{fig:complexity}).  At real-world
scale with tens of delivery points, multiple vehicles, and tight operational
time windows, exact solvers exhaust their time budgets before finding
a provably good solution, while pure heuristics leave measurable
distance savings unrealized.

Quantum and quantum-inspired computing offer potential speedup for
combinatorial optimization via Quadratic Unconstrained Binary
Optimization (QUBO) models~\cite{glover2018qubo, lucas2014ising}.  The
Quantum Approximate Optimization Algorithm (QAOA)~\cite{farhi2014qaoa}
and Simulated Bifurcation Machines (SBM)~\cite{goto2016qbm,goto2019sbm}
are two leading approaches.  SBM is quantum-inspired in the specific
sense that it descends from bifurcation-based adiabatic quantum
computation using coupled nonlinear oscillators, but replaces direct
quantum evolution by an efficiently simulated classical nonlinear
Hamiltonian dynamics.  This makes it especially attractive for dense
QUBO/Ising subproblems because all variables can be updated
simultaneously and the dominant computation is a matrix-vector
interaction over the coupling matrix~\cite{goto2019sbm,tatsumura2024sbm}.
Prior work~\cite{feld2019hybrid, harwood2021routing} has formulated small VRP instances as QUBOs, but monolithic encodings of full CVRPTW remain intractable at practical scale: qubit counts,
penalty-coefficient stability, and time-window feasibility enforcement
all degrade rapidly with instance size (see Supplementary~Information).

We address this gap with a three-stage \textbf{A}ssignment-\textbf{R}outing-\textbf{T}ime-check
(A-R-T) hybrid pipeline in which quantum and quantum-inspired solvers
handle only small per-vehicle subproblems, well within their tractable
range.  Classical MILP (solved by Gurobi ~\cite{gurobi2024}) anchors every stage: it provides
complete end-to-end routing for smaller instances and assignment-only
solutions for larger ones, with quantum and quantum-inspired solvers
contributing route-sequence refinement as an additive polishing step.
As instance size grows, the pipeline transitions naturally from a
single-solver classical approach to an orchestrated multi-stage
architecture, leveraging each solver where it is most effective.  The
framework is evaluated end-to-end on real London traffic data across
five instances (10 to 70 customers) and, to our knowledge, represents
the first multi-backend empirical study of quantum and quantum-inspired
methods on real-world CVRPTW instances at this scale.

\textbf{Contributions:}
\begin{enumerate}
\item A scalable A-R-T hybrid pipeline separating CVRPTW into classical
      MILP assignment/routing and per-vehicle Traveling Salesman Problem (TSP) QUBO polishing with a
      strict monotone-improvement guarantee.
\item A time-window-aware QUBO formulation with incumbent-scaled penalty
      calibration and progressive bias-escalation for improved
      feasible-candidate yield.
\item A rigorous multi-backend benchmark (Gurobi, Toshiba SQBM+,
      QAOA on Rigetti Ankaa-3 / IQM Emerald) on five
      real London instances from 10 to 70 customers.
\item Empirical characterization of the current QAOA hardware boundary
      for TSP-type QUBO constraint satisfaction ($m\!=\!4$, 16 qubits).
\end{enumerate}

\section{Problem Definition and Dataset}

A CVRPTW instance is defined on a directed graph $G=(N,A)$ with
$N=\{0\}\cup C$ (depot $0$, customers $C=\{1,\ldots,n\}$). The arc set A contains all feasible directed connections between nodes.
Arc $(i,j)\in A$ has travel distance $c_{ij}$ (m) and travel time $\tau_{ij}$
(min).  Each customer $i$ has demand $d_i$, service time $s_i$ (min),
and time window $[a_i, b_i]$.  All $|K|$ vehicles are homogeneous with
capacity $Q$ and operate within depot window $[a_0, b_0]$.

\textbf{Safe-arc pruning.} Arcs that can never satisfy time-window feasibility under earliest departure conditions are removed from consideration. An arc is retained only if servicing node $i$ at its earliest time, followed by travel to node $j$, can still respect the latest allowable service time at node $j$ i.e., arc $(i,j)$ is retained only if
$a_i + s_i + \tau_{ij} \le b_j$; otherwise it cannot be part of any
feasible solution. This reduces $A$ to the pruned set $A'\subseteq A$. This reduction decreases the number of decision variables and constraints without excluding any feasible solutions, thereby significantly improving solver performance.

\textbf{Benchmark instances.}  Five instances were constructed from real
London business locations with AWS real-time traffic distance/time
matrices (Table~\ref{tab:instances}).

\begin{table}[t]
\centering
\caption{London CVRPTW Benchmark Instances}
\label{tab:instances}
\renewcommand{\arraystretch}{1.05}
\begin{tabular}{@{}lcccc@{}}
\toprule
ID & $|C|$ & $|K|$ & TW (min) & MILP strategy \\
\midrule
C10K2  & 10 & 2  & 480 & Full CVRPTW MILP \\
C25K5  & 25 & 5  & 480 & Full CVRPTW MILP \\
C40K8  & 40 & 8  & 600 & Assignment MILP  \\
C55K11 & 55 & 11 & 625 & Assignment MILP  \\
C70K14 & 70 & 14 & 700 & Assignment MILP  \\
\bottomrule
\end{tabular}
\end{table}

\section{Hybrid Optimization Architecture}
\label{sec:arch}

The pipeline decomposes CVRPTW into three functionally distinct stages
named \textbf{A}-\textbf{R}-\textbf{T}: \textbf{A}ssignment
(deciding which customers are served by which vehicle),
\textbf{R}outing (sequencing each vehicle's assigned customers into a
minimum-distance tour), and \textbf{T}ime-check (validating
time-window compliance and enforcing monotone improvement).  Classical
MILP handles the A and R stages directly for smaller instances; for
larger instances it handles A while quantum and quantum-inspired solvers
refine R.  The T stage is always classical, providing a hard feasibility
and improvement gate that guarantees the final solution never regresses
below the classical baseline regardless of which backend produced the Stage~R candidate.

\subsection*{Three-Stage A-R-T Pipeline}

\textbf{Stage A — Customer-Vehicle Assignment.}
Gurobi establishes a globally feasible incumbent by solving the
assignment problem: the full CVRPTW MILP
(see Section~\ref{sec:formulation}) for $|C|\le 25$, jointly
optimising assignment and routing in a single model; or the
assignment-only MILP followed by brute-force resequencing
($m!$ permutations, earliest-departure simulation, for $m\le 8$) for
larger instances.  All five instances reach a low assignment-gap,
providing a reliable customer-to-vehicle partition as input to Stage~R.

\textbf{Stage R — Per-Vehicle Routing.}
Given the Stage~A assignment, each vehicle's customer subset $C_k$ is
sequenced independently as a TSP subproblem
(see Section~\ref{sec:formulation}).  For the classical baseline,
brute-force or MILP resequencing is used; for quantum polishing, each
$C_k$ is encoded as a QUBO and submitted in parallel to SQBM+ and
QAOA.  In practice, $|C_k|=2$--$6$ customers per vehicle
($m^2=4$--$36$ QUBO variables), keeping every subproblem within the
tractable range of both SQBM+ and current QAOA hardware.
Monolithic CVRPTW-QUBO alternatives were evaluated and found impractical
even for C10K2; see Supplementary~Information for experimental evidence.

\textbf{Stage T — Time-Check and Acceptance Gate.}
Each candidate route returned by Stage~R is evaluated in order:
(i)~arc-level time-window pre-screen; (ii)~forward time propagation
accounting for travel time, service duration, and waiting time;
(iii)~strict improvement test — accepted if and only if the route is
time-window feasible \emph{and} strictly shorter than the current
incumbent.  Failed candidates trigger a \texttt{keep\_original}
fallback to the Stage~A route.  This gate guarantees
\emph{monotone improvement}: the final solution is always at least as
good as the classical baseline, regardless of which backend produced
the Stage~R candidate.

\section{Problem Formulation}
\label{sec:formulation}

\subsection{Full CVRPTW MILP}

For smaller problem instances, CVRPTW is formulated as a full Mixed Integer Linear Program (MILP). In this formulation, customer-to-vehicle assignments, routing decisions, and time-window feasibility are jointly optimized within a single integrated model. The formulation uses the following variables: $x_{ijk}\!\in\!\{0,1\}$  {1 if vehicle $k$ travels directly from node $i$ to node $j$},
$y_{ik}\!\in\!\{0,1\}$ {1 if customer $i$ is assigned to vehicle $k$},
$v_k\!\in\!\{0,1\}$ {1 if vehicle $k$ is used (leaves and returns to depot)}, and
$t_{ik}\ge 0$  {service start time of vehicle $k$ at node $i$}.

The objective minimizes total travel distance of entire fleet:
\begin{equation}
\min \;\sum_{k\in K}\sum_{(i,j)\in A'} c_{ij}\,x_{ijk}.
\label{eq:milp_obj}
\end{equation}

Key constraint groups enforce customer coverage
($\sum_k y_{ik}=1$), capacity ($\sum_i d_i y_{ik}\le Q$), flow
conservation, Big-$M$ time propagation, and time-window bounds.
The full enumeration with parameter definitions is given in
Supplementary~Information.

The model scales as $\mathcal{O}(|K|\cdot|C|^2)$ in both variables and
constraints: 13,776 variables and approximately 14,500 constraints for
40 customers (C40K8 with $|K|{=}8$) motivating the decomposed strategy for larger instances (Section~\ref{sec:arch}).

Gurobi receives this formulation in MPS format via the Strangeworks
cloud layer~\cite{strangeworks2024} (a quantum cloud platform that
provides unified API access to QPU and classical, quantum-inspired and quantum compute backends
including Gurobi, Toshiba SQBM+, Rigetti, and IQM); see Supplementary~Information
for quantization and Big-$M$ construction details.

\subsection{Per-Vehicle TSP QUBO Formulation}
\label{sec:qubo}

The vehicle-local routing problem is formulated as a soft, time-window–aware Traveling Salesman Problem (TSP) for each vehicle. Given a fixed vehicle $k$, let $C_k = \{i_1, \dots, i_m\}$ denote the set of customers assigned to that vehicle, the objective is to determine an ordering of these customers that minimizes the total travel distance. This sequencing problem is encoded as a QUBO model using position-based assignment variables. Specifically, binary variables
\[
z_{i,p} \in \{0,1\}, \qquad i \in C_k,\; p \in \{1,\dots,m\},
\]
are introduced, where $z_{i,p} = 1$ indicates that customer $i$ is visited in position $p$ of the route.

The complete QUBO is represented internally as a Binary Quadratic
Model (BQM), a standard abstraction over binary variables with
linear biases $h_a$ (diagonal of the coupling matrix $\mathbf{Q}$)
and quadratic interactions $J_{ab}$ (off-diagonal entries).  This BQM
representation serves as the common interface: it is submitted
directly to Toshiba SQBM+
(Section~\ref{sec:sqbm_submission}) and, after translation to an Ising
Hamiltonian, to gate-based QAOA backends
(Section~\ref{sec:qubo_to_ising}).

The goal of the optimization is twofold: to construct routes that are short in terms of total travel distance, while also discouraging sequences that are likely to violate time-window constraints. Since hard time-window constraints are not explicitly enforced in the quadratic formulation, feasibility is instead guided through a penalty mechanism embedded in the objective function. To achieve this, the QUBO formulation uses an objective-only, vehicle-local distance matrix that augments the original travel cost with a time-window–aware penalty. This modified distance captures both the physical travel cost and an estimate of the risk of arriving late at subsequent customers. The effective vehicle-local distance matrix is defined as
\[
D^{(k)}_{ij}
=
c_{ij}
+
\lambda_{\mathrm{TW}}^{\mathrm{route}}
\cdot
\max\bigl(0,\;\hat{t}^{\mathrm{lb}}_i + s_i + \tau_{ij} - b_j\bigr),
\]
where the time-window penalty term $\lambda_{\mathrm{TW}}^{\mathrm{route}}$ increases the cost of transitions that are likely to induce time-window violations. This formulation does not enforce feasibility but biases the optimization away from high-risk sequencing decisions. Here $\hat{t}^{\mathrm{lb}}_i$ is the optimistic lower bound on the earliest possible service start at node $i$, computed using depot reachability.  The $\max$ term measures how much the earliest feasible departure from ${i}$ overshoots ${j}$'s latest service deadline, penalizing likely-infeasible arcs without hard exclusion. The weight $\lambda_{\mathrm{TW}}^{\mathrm{route}}=0.10 \times P_{\mathrm{inc}}$ uses an incumbent-scaled penalty $P_{\mathrm{inc}}= \alpha \cdot d^{(k)}_{inc} (\alpha=10\)), where $d^{(k)}_{inc}$ is Gurobi's current best route distance for vehicle $k$ from \eqref{eq:milp_obj}.  This ties the penalty magnitude to the actual routing scale of each vehicle: a 1-unit time-window violation costs one full incumbent route length, making infeasible arcs strongly unattractive while preserving the routing signal. As Gurobi finds shorter routes, $P$ contracts automatically, tightening the penalty progressively without any per-instance tuning.

Using this modified distance metric, the vehicle-local routing problem is expressed as a quadratic objective over position assignment variables:
\begin{equation}
\min\;
\sum_{i\in C_k} D^{(k)}_{0i} z_{i,1}
+
\sum_{p=1}^{m-1}\sum_{i\neq j\in C_k}
D^{(k)}_{ij} z_{i,p}z_{j,p+1}
+
\sum_{i\in C_k} D^{(k)}_{i0} z_{i,m}.
\label{eq:tsp_obj}
\end{equation}

The three terms in \eqref{eq:tsp_obj} cover depot-to-first, consecutive-pair, and last-to-depot transitions respectively.

Permutation constraints are added as quadratic penalties with weight $P_{\mathrm{perm}}$ to \eqref{eq:tsp_obj}:
\begin{equation}
+P_{\mathrm{perm}}\!\left[\sum_{p=1}^m\!\Bigl(\textstyle\sum_{i\in C_k}z_{i,p}-1\Bigr)^{\!2}
          +\sum_{i\in C_k}\!\Bigl(\textstyle\sum_{p=1}^mz_{i,p}-1\Bigr)^{\!2}\right].
\label{eq:qubo_perm}
\end{equation}
Together these two terms guarantee a complete, conflict-free permutation of $C_k$.

The QUBO is encoded as a symmetric $m^2 \times m^2$ coupling matrix $\mathbf{Q}$, whose $\binom{m^2}{2}$ off-diagonal entries represent pairwise variable interactions among the $m^2$ binary position variables: $9$ variables and $36$ couplings for $m=3$; $36$ variables and $\sim630$ couplings for $m=6$. This $\mathcal O(m^4)$ growth in couplings is the core hardware bottleneck: physical quantum and quantum-inspired machines must resolve all couplings simultaneously, and limited qubit connectivity inflates embedding cost rapidly with $m$. The routing objective populates $\mathbf{Q}$ with consecutive-position couplings ($z_{i,p}$–$z_{j,p+1}$), while the permutation penalties add within-position and within-customer couplings, densely filling $\mathbf{Q}$.

\subsubsection{From Quantum Bifurcation Machines to SQBM+}
\label{sec:qbm_to_sqbm}
Simulated Bifurcation Machines originate from the quantum bifurcation
machine (QBM) model proposed by Goto, in which a network of
Kerr-nonlinear parametric oscillators is coupled according to an Ising
cost function and evolved adiabatically through bifurcation points.
In that quantum model, the two stable oscillation branches encode the
two Ising spin values, and the signs of the final oscillator amplitudes
represent a candidate spin configuration~\cite{goto2016qbm}.  The later
simulated bifurcation (SB) algorithm replaces direct quantum evolution
with a classical approximation of the nonlinear oscillator network and
numerically simulates Hamiltonian dynamics exhibiting bifurcation
phenomena~\cite{goto2019sbm}.  Toshiba's SQBM+ service is an
implementation of this SB family of quantum-inspired algorithms for
quadratic discrete optimization~\cite{toshiba2023sqbmTips,tatsumura2024sbm}.

For a QUBO objective
\begin{equation}
E_{\mathrm{QUBO}}(z)=\sum_{a<b}Q_{ab}z_a z_b+\sum_a Q_{aa}z_a,
\label{eq:qubo_energy}
\end{equation}
the binary variables $z_a\in\{0,1\}$ are mapped to classical Ising
spins $s_a\in\{-1,+1\}$ via $z_a=(1-s_a)/2$, yielding the classical
Ising energy:
\begin{equation}
E_{\mathrm{Ising}}(s)=\sum_{a<b}J_{ab}\,s_a s_b+\sum_a h_a\,s_a,
\label{eq:ising_sqbm}
\end{equation}
The constant offset is suppressed as it does not affect minimisation.
The Ising coefficients are derived from $\mathbf{Q}$:
\begin{equation}
J_{ab}=\tfrac{1}{4}Q_{ab},\qquad
h_a=-\tfrac{1}{2}Q_{aa}-\tfrac{1}{4}\!\sum_{b\neq a}Q_{ab}.
\label{eq:ising_coeffs}
\end{equation}
SQBM+ minimises~\eqref{eq:ising_sqbm} over classical $s_a$ by evolving
continuous position--momentum variables associated with the spins.

Intuitively, as the bifurcation parameter is swept, the
energy landscape changes from a single basin into many minima
corresponding to discrete spin configurations.  The SB dynamics combines
an adiabatic component, which tracks emerging low-energy minima, with
an ergodic or chaotic-search component, which increases the probability
of visiting better minima in the allowable phase-space region
~\cite{goto2019sbm,tatsumura2024sbm}.  This differs from gate-based
QAOA: SQBM+ does not construct a quantum circuit, does not consume
qubits, and does not require the QUBO-to-Ising Hamiltonian to be
compiled into two-qubit gates.  Instead, it accepts the dense QUBO/BQM
couplings directly and exploits the high parallelism of the SB update.

\subsubsection{SQBM+ Submission and Bias Escalation}
\label{sec:sqbm_submission}
To combine exploration and exploitation,
three sequential calls are made to SQBM+ service per vehicle.  Each call extends the
base BQM with two soft-bias terms: negative linear coefficients on
$z_{i,p}$ variables matching the incumbent route
(\emph{incumbent position bias}), and reduced costs on the first- and
last-position variables (\emph{depot-edge bias}).  Bias strengths
escalate across attempts; candidate pools from all three calls are
merged before the Stage~T acceptance gate
(Section~\ref{sec:arch}):
\begin{center}\small
\begin{tabular}{@{}ccc@{}}
\toprule
Attempt & $\lambda_{\mathrm{bias}}$ & $\lambda_{\mathrm{depot}}$ \\
\midrule
1 & $0.05\,P_{\mathrm{perm}}$ & $0.06\,P_{\mathrm{perm}}$ \\
2 & $0.10\,P_{\mathrm{perm}}$ & $0.08\,P_{\mathrm{perm}}$ \\
3 & $0.15\,P_{\mathrm{perm}}$ & $0.10\,P_{\mathrm{perm}}$ \\
\bottomrule
\end{tabular}
\end{center}
Attempt~1 applies a low initial bias for broad exploration, and later
attempts progressively tighten the bias to exploit the incumbent
neighbourhood.

\subsubsection{QUBO-to-Ising Translation for Gate-Based Hardware}
\label{sec:qubo_to_ising}

Gate-based backends encode the same BQM as a \emph{quantum} Ising
Hamiltonian operator.  Each binary variable $z_{i,p}$ is replaced by
the operator $\tfrac{1}{2}(I-Z_{i,p})$, the quantum analogue of the
classical substitution $z_a\!=\!(1-s_a)/2$ in~\eqref{eq:ising_sqbm}.
This yields the same real-valued coefficients $J_{ab}$, $h_a$
as in~\eqref{eq:ising_coeffs}; the constant offset $C\cdot I$
shifts all eigenvalues equally and is therefore dropped:
\begin{equation}
H \;=\; \sum_{a<b} J_{ab}\,Z_a Z_b \;+\; \sum_a h_a\,Z_a,
\label{eq:ising}
\end{equation}
where $Z_a$ denotes the Pauli-$Z$ operator on qubit $a$
in the flattened register. The translation is performed via \texttt{bqm\_to\_ising()} in the
Amazon Braket SDK~\cite{braket2024}, with coefficients rescaled to
hardware-compatible ranges.

\subsubsection{QAOA Circuit Construction}
The QAOA ansatz alternates a cost unitary $e^{-i\gamma H}$, where $H$
is the route-cost Hamiltonian~\eqref{eq:ising}, with a mixer unitary
$e^{-i\beta B}$, where $B\!=\!\sum_a X_a$ is the transverse-field
operator ($X_a$ is the Pauli-$X$ on qubit $a$). $B$
decomposes as a tensor product, $e^{-i\beta B}\!=\!\bigotimes_a
\mathrm{RX}(2\beta)_a$, a single-qubit layer.
The circuit uses $p$ layers; $\gamma$ and $\beta$ are variational
parameters calibrated per backend.
Each binary variable $z_{i,p}$ maps to a unique qubit via the
position-major index $\mathit{vid}=p\cdot m+i+1$, which flattens
the $m\times m$ variable grid into a 1-based qubit register of
length $m^2$ and groups same-position variables contiguously to
reduce two-qubit gate distance.

\subsubsection{Candidate Validation and Diagnostics}
Every candidate route returned by the QUBO solver undergoes strict CVRPTW validation as described in Stage~T (Section~\ref{sec:arch}):
capacity feasibility, forward time propagation, and time-window
compliance per customer, with acceptance conditional on strict distance
improvement over the incumbent.  Infeasible or non-improving candidates
are rejected from the solution but retained for diagnostics, enabling
analysis of the magnitude and structure of time-window violations
at the vehicle level.

\subsection{Assignment-Only MILP for Large Instances}

For $|C|>25$, the full MILP is intractable within practical time
limits, so Stage~A decomposes into two steps: an
\emph{assignment-only} MILP that partitions customers across vehicles,
followed by independent per-vehicle sequencing in Stage~R.  Vehicle
\emph{anchors} are computed by running $k$-means with $k=|K|$ on
customer coordinates, yielding $|K|$ centroids that serve as proxy
territory representatives.  The proxy assignment cost
$c_{ik}$ is the Euclidean distance from customer $i$ to the centroid
of vehicle $k$'s cluster, encouraging geographically compact groupings
without modelling routing arcs explicitly.  The binary variables are:
$y_{ik}$ and $v_k$ as defined in the full CVRPTW formulation, and auxiliary
$w_{ijk}\!\in\!\{0,1\}$ for selected high-risk customer pairs
$(i,j)$ (linearising the conjunction ``$i$ and $j$ both on $k$''). $w_{ijk}$ equals 1 if and only if both customer $i$ and customer $j$ are co-assigned to vehicle $k$ (i.e., $y_{ik}=1$ and $y_{jk}=1$ simultaneously).

Cardinality bounds keep each vehicle's subproblem tractable for the subsequent
Stage~R:
\begin{equation}
\underline{M}\,v_k \;\le\; \sum_{i\in C} y_{ik} \;\le\; \overline{M}\,v_k
\qquad\forall k\in K,
\label{eq:cardinality}
\end{equation}
with $\underline{M}=2$, $\overline{M}=6$ in the current
implementation.  Assignment-to-usage linking is enforced by
$y_{ik}\le v_k\ \forall i,k$, and all vehicles are forced active
($v_k=1$) so that customers are spread across the full fleet.

Because time windows are not modelled explicitly, a soft penalty
discourages assigning strongly incompatible pairs to the same vehicle.
The time-window non-overlap deficit for pair $(i,j)$ is:
\begin{equation}
\Delta_{ij}^{\mathrm{TW}}=\max\!\bigl(0,\;\max(a_i,a_j)-\min(b_i,b_j)\bigr),
\label{eq:tw_deficit}
\end{equation}
where $\Delta_{ij}^{\mathrm{TW}}=0$ indicates compatible windows and
larger values signal greater incompatibility.  Only pairs whose deficit
exceeds a threshold are retained, capped at $|\mathcal{R}|\le 100$
highest-risk pairs.  For each $(i,j)\in\mathcal{R}$ and vehicle $k$,
the auxiliary $w_{ijk}$ linearises the co-assignment conjunction via:
$w_{ijk}\le y_{ik}$,\ $w_{ijk}\le y_{jk}$,\
$w_{ijk}\ge y_{ik}+y_{jk}-1$.

The full assignment objective is:
\begin{equation}
\min \;\sum_{i,k} c_{ik}\,y_{ik}
  + \lambda_{\mathrm{use}}\!\sum_k v_k
  + \!\!\sum_{(i,j)\in\mathcal{R}}\sum_k
    \lambda_{\mathrm{TW}}^{\mathrm{assign}}\,\Delta_{ij}^{\mathrm{TW}}\,w_{ijk},
\label{eq:assign_obj}
\end{equation}
where the three terms minimise anchor-based clustering cost, penalise
unnecessary vehicle activation, and soft-penalise time-window-incompatible
co-assignments respectively.  Routing and time-window feasibility are
deferred entirely to Stage~R and Stage~T. This separation reduces the model dramatically: for C70K14, the
assignment MILP yields 994 binary variables and 1,106 constraints
versus 13,776 variables for the equivalent full MILP.
The scaling extends structurally beyond C70K14: a C100K20 instance
would require approximately 1,400 assignment variables—still
millisecond-solvable—while the cardinality cap $\overline{M}=6$
in~\eqref{eq:cardinality} ensures every per-vehicle QUBO remains
bounded at $m^2\!\le\!36$ qubits irrespective of total customer
count or fleet size.  Stage~T is vehicle-local and trivially
parallelisable.  The pipeline's computational envelope is therefore
set by per-vehicle subproblem size, not by instance scale; the
binding constraint at larger instances is assignment-stage
time-window tightness, not solver computability.

\begin{figure}[t]
  \centering
  \includegraphics[width=\columnwidth]%
    {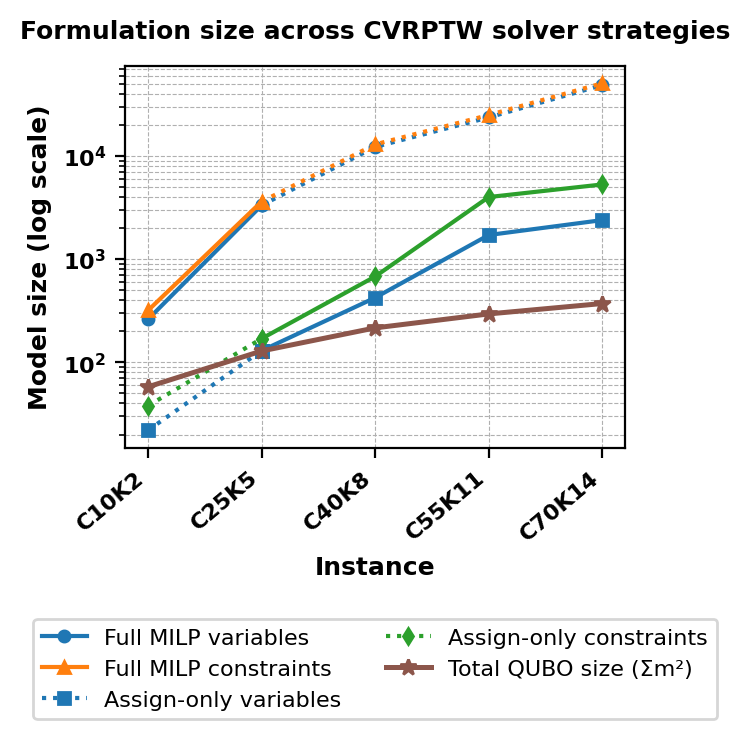}
  \caption{Model formulation size (variables and constraints, log scale)
    across the three solver strategies used in the A-R-T pipeline.
    Full CVRPTW MILP grows as $\mathcal{O}(|K|\cdot|C|^2)$ in both
    variables and constraints; the assignment-only MILP reduces to
    $\mathcal{O}(|K|\cdot|C|)$; total QUBO size ($\sum_k m_k^2$,
    summed across all per-vehicle subproblems) scales as
    $\mathcal{O}(m^2)$ per vehicle, remaining tractable even at
    70 customers.  The cardinality cap $\overline{M}=6$
    bounds the per-vehicle QUBO at 36 qubits regardless of instance
    size, so the three curves retain the same ordering beyond C70K14.}
  \label{fig:complexity}
\end{figure}

\section{Implementation}
\label{sec:impl}

\subsection{Gurobi Configuration}

Input data are integer-quantized (distances in metres, times in
minutes) before model construction, ensuring numerical consistency
with downstream QUBO formulations.  Models are exported to MPS format
and submitted via the Strangeworks cloud execution layer~\cite{strangeworks2024}; a
cached-solution toggle allows fast feasibility iteration without
re-solving.

Smaller instances (C10K2, C25K5) are solved as a full CVRPTW MILP
with parameters tuned for solution quality.
Larger instances use an assignment-only MILP; the best
assignment partition seeds Stage~R QUBO subproblems.
Table~\ref{tab:gurobi} summarizes Gurobi solve costs and outcomes.

\begin{table}[t]
\centering
\caption{Gurobi Solve Summary}
\label{tab:gurobi}
\renewcommand{\arraystretch}{1.05}
\begin{tabular}{@{}llrrc@{}}
\toprule
ID & Strategy & Time (s) & Cost (\$) \\
\midrule
C10K2  & Full CVRPTW MILP     & 60 & 0.056 \\
C25K5  & Full CVRPTW MILP     & 60  & 0.17 \\
C40K8  & Assignment MILP      & $<$0.1 & 3.29 \\
C55K11 & Assignment MILP      & $<$0.1 & 0.0002 \\
C70K14 & Assignment MILP      & $<$0.1 & 0.0002 \\
\bottomrule
\end{tabular}
\end{table}

\subsection{Toshiba SQBM+ Configuration}

All QUBO inputs are integer-valued (mandatory for SQBM+ numerical
stability): distances in metres and times in minutes are rounded to
integers before BQM construction.  Vehicle subproblems are dispatched
independently and can run in parallel; each call targets a single
vehicle's $m^2$-variable QUBO.

Each per-vehicle QUBO is submitted via the SQBM+ REST API with
\texttt{steps=0}, \texttt{loops=50}, \texttt{timeout=30}\,s, and
\texttt{maxout=24}, held constant across all
instance sizes so that outcome differences reflect problem structure
rather than solver retuning~\cite{toshiba2023sqbmTips}.
\texttt{steps=0} enables automatic step adaptation: the API
adjusts the number of steps each time it finds a
solution~\cite{toshiba2023sqbmTips}, avoiding a fixed iteration
budget.  A \emph{loop} is a batch of repeated SB runs, each
producing one candidate; \texttt{loops=50} bounds the total search
width per request.  \texttt{timeout=30}\,s sets the wall-clock
limit (the API default is 10~s).  \texttt{maxout=24} returns up to
24 low-energy candidates ranked in descending objective-value
order~\cite{toshiba2023sqbmTips}, maximising the pool presented to
Stage~T.  The three-attempt
bias escalation schedule is described in
Section~\ref{sec:sqbm_submission}.  If no attempt yields a feasible
improving candidate, a \texttt{keep\_original} fallback retains the
Stage~A route unchanged.  Table~\ref{tab:sqbm} summarises SQBM+
execution parameters.

\begin{table}[t]
\centering
\caption{SQBM+ Execution Summary}
\label{tab:sqbm}
\renewcommand{\arraystretch}{1.05}
\begin{tabular}{@{}lccrr@{}}
\toprule
ID & Time (s) & Cost (\$) & Net result \\
\midrule
C10K2 & 2.0  & 1.02 & $+0.00\%$ \\
C25K5 & 1.8  & 1.02 & $\mathbf{-2.80\%}$ \\
C40K8 & 0.8  & 3.75 & $+0.00\%$ \\
C55K11 & 1.1  & 0.63 & $\mathbf{-0.22\%}$ \\
C70K14 &  1.4  & 0.80 & $+0.00\%$ \\
\bottomrule
\end{tabular}
\end{table}

\subsection{Quantum QAOA Configuration}

QAOA circuits are submitted via the Amazon Braket SDK~\cite{braket2024}
(AWS's fully managed quantum computing service providing a common SDK
and cloud access to gate-based QPUs and state-vector simulators).
Two QPU backends are used in production.
Rigetti Ankaa-3 (\texttt{us-west-1}, superconducting, 79~qubits,
native gate set $\{\mathrm{RZ},\mathrm{RX},\mathrm{ISWAP}\}$) handles
instances C10K2 and C25K5.
IQM Emerald (\texttt{eu-north-1}, superconducting, native gate set
$\{\mathrm{PRX},\mathrm{CZ}\}$, where
$\mathrm{PRX}(\theta,\varphi)\!=\!\mathrm{RZ}(-\varphi)\cdot\mathrm{RX}(\theta)\cdot\mathrm{RZ}(\varphi)$
is the phased-RX rotation) handles C40K8
through C70K14.
Backend availability is discovered at runtime via
Strangeworks~\cite{strangeworks2024}.
The SV1 state-vector simulator (up to 34~qubits) is the preferred
fallback when a QPU is unavailable or queue wait exceeds 1,800~s.
Job completion is polled at 2-s intervals up to that timeout.
A hard per-vehicle cap of $m\!\le\!6$ customers (36~qubits) is enforced
before circuit submission.
Larger subproblems fall back to the Stage~A route without a QAOA call.

The QAOA parameters $\gamma$ and $\beta$ have distinct physical roles.
$\gamma$ controls the phase rotation applied to each computational basis
state proportional to its value under $H$~\eqref{eq:ising}: larger
$\gamma$ sharpens the cost landscape, giving high-quality (low-cost)
routes a stronger phase signal relative to poor ones.
$\beta$ controls the mixing angle of the mixer unitary $e^{-i\beta B}$:
it governs how broadly the circuit explores
the space of route permutations between cost evaluations.
Circuit depth and calibration are tuned per backend.

In both cases, $(\gamma,\beta)$ pairs are evaluated on the Amazon Braket
SV1 state-vector simulator: for each candidate pair a QAOA circuit is
executed on SV1, the resulting measurement distribution is decoded into
candidate tours, and each tour is assessed by route distance and full
CVRPTW feasibility.
The pair yielding the best feasible routing solution is selected for
the QPU run; route quality and feasibility take precedence over minimum
expected Ising energy, since the operational objective is routing cost
rather than Hamiltonian energy.
The same procedure applies to QPU executions, with the execution
backend as the only difference.

\textit{Ankaa-3} uses $p=1$ layers for C10K2 and C25K5.
Parameters $(\gamma^*,\beta^*)=(1.70,\,0.0975)$ are found via an offline
SV1 grid search on a representative subproblem and fixed globally across
all vehicles; no per-vehicle sweep is performed because the shallow
$p\!=\!1$ landscape is stable across vehicle sizes at this instance scale.
Production runs use \textbf{50 shots} per vehicle.

\textit{Emerald} uses $p=1$ layers for C40K8 and $p=2$ layers for
C55K11 and C70K14.
For C40K8, a single fixed pair $(\gamma^*,\beta^*)\!=\!(1.70,\,0.0975)$
is applied uniformly to all vehicles (same as Ankaa-3; no per-vehicle
sweep is performed for this instance).
For C55K11 and C70K14, parameters are selected per vehicle via an SV1
grid search.
The sweep consistently selects $(\gamma^*,\beta^*)\!=\!(2.0,\,0.08)$
for vehicles with $m\!\le\!5$ customers; vehicles at the maximum size
$m\!=\!6$ use $(\gamma^*,\beta^*)\!=\!(1.2,\,0.40)$.
For $p=2$ circuits, the two-layer angles are constructed from the
selected pair as $\boldsymbol{\gamma}=(\gamma^*/2,\,\gamma^*)$ and
$\boldsymbol{\beta}=(\beta^*,\,\beta^*/2)$, giving
$\boldsymbol{\gamma}=(1.0,2.0)$, $\boldsymbol{\beta}=(0.08,0.04)$
for smaller vehicles and
$\boldsymbol{\gamma}=(0.6,1.2)$, $\boldsymbol{\beta}=(0.40,0.20)$
for 6-customer vehicles.
Production runs use \textbf{100 shots} per vehicle for C40K8 and
C70K14, and \textbf{200 shots} for C55K11.

All QPU runs use a fallback decoder because high-probability QAOA
bitstrings frequently violate the permutation constraints.  Bitstrings
from the shot histogram are ranked by count (descending); a bitstring is
a \emph{valid permutation} if and only if the reshaped $m\times m$
assignment matrix $Z$ satisfies
$\sigma=\sum_p|\sum_i Z[i,p]-1|+\sum_i|\sum_p Z[i,p]-1|=0$
(one customer per position and one position per customer).  The first
permutation-valid state in the ranked list is accepted as the candidate
route; if none of the top-20 states satisfies $\sigma=0$, the Stage~A
route is retained (\texttt{keep\_original}).  Across all production QPU
runs, the rank-1 (most-frequent) bitstring never achieved $\sigma=0$;
all accepted routes were obtained from lower-ranked states, highlighting
the difficulty of enforcing permutation constraints on current
gate-based hardware. The position-based TSP encoding requires $m^2$ logical qubits
per vehicle ($m$ customers × $m$ route positions), yielding
4 qubits for $m=2$, 9 qubits for $m=3$, and 16 qubits for
$m=4$. Vehicle subproblems with $m\ge5$ (25+ qubits) are not
submitted to QAOA and instead retain the Stage~A route. Table~\ref{tab:qaoa} summarizes QAOA results.

\begin{table}[t]
\centering
\caption{QAOA Execution Summary}
\label{tab:qaoa}
\renewcommand{\arraystretch}{1.05}
\begin{tabular}{@{}llcrr@{}}
\toprule
ID & Backend & Time (s) & Cost (\$) & Net result \\
\midrule
C10K2  & Ankaa-3 & 8.9  & 0.69  & $+0.00\%$ \\
C25K5  & Ankaa-3 & 21.4 & 1.89 & $+24.63\%$ \\
C40K8  & Emerald & 32.0 & 6.05 & $+2.58\%$ \\
C55K11 & Emerald & 45.2 & 3.80 & $+0.00\%$ \\
C70K14 & Emerald & 56.1  & 4.83 & $+0.00\%$ \\
\bottomrule
\end{tabular}
\end{table}

\subsection{Run Management and Validation Pipeline}

Each solver run produces a canonical output bundle:
\texttt{jsonsolution.json} (raw solution variables),
\texttt{routes\_summary.json} (per-vehicle distance, travel time,
service time, wait time, load utilisation), and
\texttt{feasibility\_report.json} (per-constraint violation audit).
A runs index (\texttt{runs\_index.json}) maintains a newest-first
registry of all attempts per instance, enabling reproducibility and
comparison across solver configurations.

Route feasibility is verified by an independent validation pass
separate from the solver: visit-once check, depot departure and
return, capacity, arc feasibility against $A'$, and full forward
time propagation with departure/arrival/wait decomposition per arc.
This separation ensures that solver-reported feasibility is never
trusted without independent confirmation, and that infeasible
candidates from SQBM+ and QAOA are correctly diagnosed rather than
silently discarded.

\section{Experimental Results}
\label{sec:results}

\begin{figure*}[!t]
  \centering
  \includegraphics[width=\textwidth]{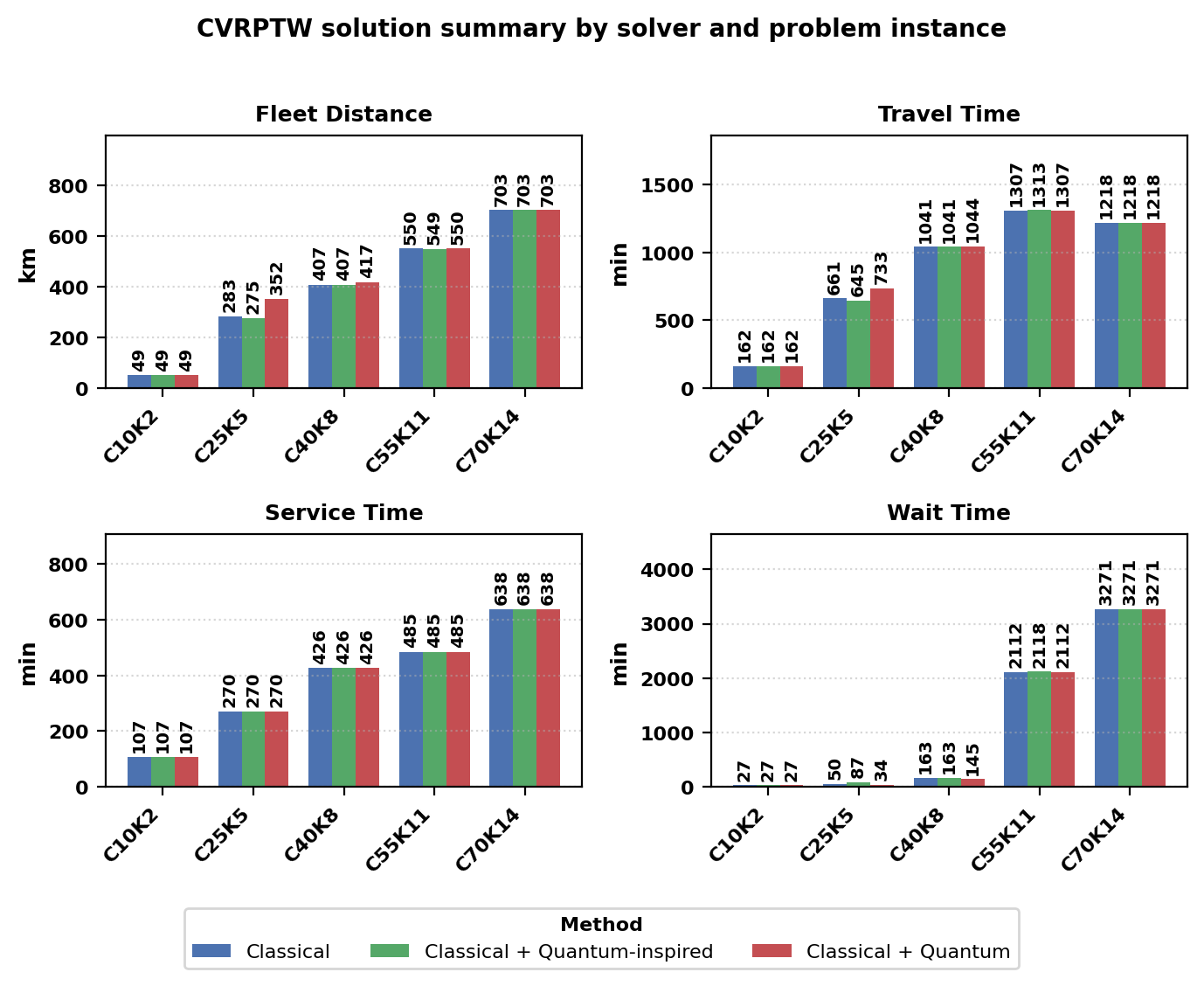}
  \caption{CVRPTW solution quality across all five instances and three
    solver configurations: Classical/Gurobi (blue),
    Classical+Quantum-inspired/SQBM+ (green),
    Classical+Quantum/QAOA (red).  Panels show total fleet distance
    (km), travel time (min), service time (min), and wait time (min).
    Service time is demand-determined and invariant across methods;
    differences arise in fleet distance and wait time.}
  \label{fig:summary}
\end{figure*}

\begin{figure}[t]
  \centering
  \includegraphics[width=\columnwidth, height=0.45\textheight,
    keepaspectratio]{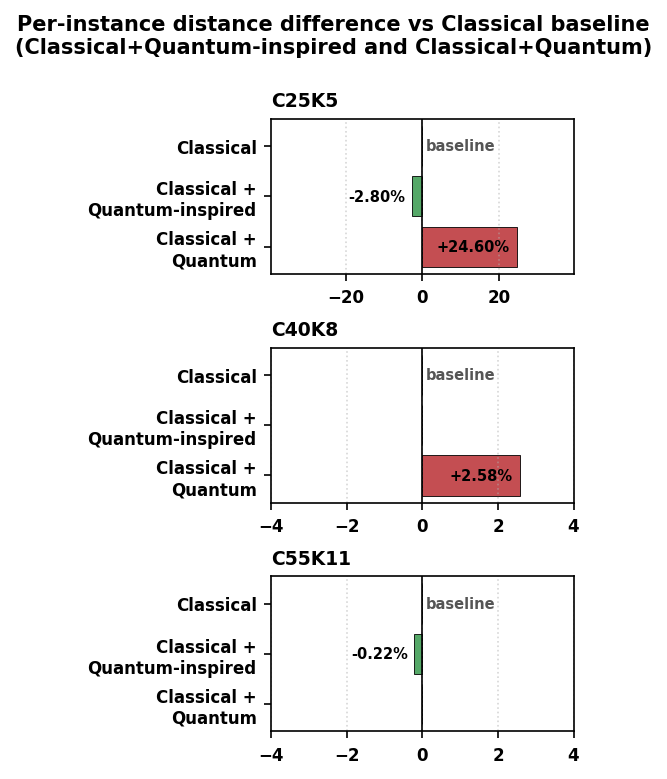}
  \caption{Distance difference vs.\ classical baseline (\%) per instance
    (C10K2 and C70K14 excluded: both hybrid methods match the classical
    baseline exactly for those instances).
    Negative = shorter/better.  SQBM+ achieves reductions for C25K5
    ($-2.80\%$) and C55K11 ($-0.22\%$).  QAOA degrades for C25K5
    ($+24.63\%$) and C40K8 ($+2.58\%$).  Remaining shown instances
    retain the classical baseline exactly under both hybrid methods.}
  \label{fig:diff}
\end{figure}

Fig.\,\ref{fig:summary} provides an overview of solution quality across
all instances and solver configurations.
Fig.\,\ref{fig:diff} shows per-instance percentage distance differences
relative to the Gurobi baseline.

\noindent\textbf{C10K2.}
Gurobi certifies global optimality at 49.4~km.
Vehicle subproblems leave negligible permutation
slack.  Both SQBM+ and QAOA confirm the classical solution.

\noindent\textbf{C25K5.}
Gurobi baseline: 282.8~km.  SQBM+ identifies improving resequencings
for two vehicles: Vehicle~2 saves 7,775~m and Vehicle~5 saves 152~m,
reducing total fleet distance to \textbf{274.9~km} ($\mathbf{-2.80\%}$).  QAOA, by contrast, produces feasible-but-suboptimal
routes for Vehicle~3 ($+60{,}951$~m) and Vehicle~4 ($+8{,}708$~m), increasing the total fleet distance to \textbf{352.5~km}
($\mathbf{+24.63\%}$).  The degradation for
Vehicles~3 and~4 reflects QAOA's failure mode at $m=4$ (16 qubits):
permutation constraints are satisfied, but the energy landscape is
too shallow to distinguish geometrically poor from geometrically
optimal permutations. Vehicles~1, and 2 return infeasible QAOA which are correctly rejected by the validation gate.  Vehicle~5 ($m=6$) returns a valid permutation
identical to the classical route ($\Delta=0$).

\noindent\textbf{C40K8.}
Assignment-only MILP baseline: 406.9~km.  Most SQBM+ candidates provide the same routes as those of Gurobi's assignment. SQBM+
diagnostic logs record best-infeasible savings of up to $13{,}762$~m
for Vehicle~1 in C70K14, confirming that assignment tightness, not QUBO
solver capability, is the limiting factor. QAOA produces feasible routes
for two vehicles, but both are longer: Vehicle~1 ($m=2$, $+909$~m) and
Vehicle~4 ($m=6$, $+9{,}599$~m), yielding $\mathbf{2.58\%}$ increase in total fleet distance.

\noindent\textbf{C55K11.}
Assignment-only MILP baseline: 550.4~km.  SQBM+ finds improving resequencings for
two vehicles: Vehicle~2 saves 730~m and Vehicle~5 saves 456~m, reducing
total to \textbf{549.2~km} ($\mathbf{0.22\%}$). QAOA (IQM Emerald) finds no net improvement ($+0.00\%$).
Permutation diagnostics reveal a size-dependent pattern: at $m=2$
(4~qubits, V10) the permutation constraint score $\sigma=0$
(valid permutation decoded), but the only non-trivial 2-customer
reordering is longer than the baseline, so Stage~T returns
\texttt{keep\_original}; at $m=3$ ($\sigma\le2$) near-valid
permutations are found but CVRPTW time-window constraints block
acceptance; at $m\ge4$ ($\sigma=4$--$12$) constraint violations
increase with subproblem size.  On Rigetti Ankaa-3 (C25K5), $m=4$
vehicles produce feasible-but-suboptimal decoded routes ($\sigma=16$),
indicating a different failure mode from Emerald: Ankaa-3 satisfies
the constraint at $m=4$ but cannot resolve the cost landscape;
Emerald violates the constraint at $m\ge4$ entirely.  This backend
-dependent expressibility gap is quantified in
Supplementary~Information~\cite{harwood2021routing}.

\noindent\textbf{C70K14.}
Assignment-only MILP baseline: 703.3~km.  Neither SQBM+ nor QAOA finds any
net improvement; both retain the classical solution exactly.
SQBM+ infeasibility diagnostics confirm improvements of $13{,}762$~m
(V1) and $7{,}140$~m (V8) were available but blocked by single
time-window node violations in the fixed assignment.
QAOA (IQM Emerald) decoded valid permutations ($\sigma=0$) for V10 and
V11 ($m=3$, 9~qubits): V11's candidate matches the baseline distance;
V10's candidate is $\mathbf{2{,}154}$~m shorter than the
baseline (21{,}138~m vs.\ 23{,}292~m, $-9.3\%$ at vehicle level),
yet Stage~T rejects it for CVRPTW time-window infeasibility in the
fixed Stage~A assignment.  This is the sharpest single-vehicle
confirmation that assignment tightness, not QUBO expressibility, is
the binding constraint: the quantum circuit found the geometrically
better route; the upstream assignment structure made it unreachable.

Table~\ref{tab:perveh} provides the complete set of non-zero per-vehicle
route changes across all instances, confirming that improvements are
sparse, vehicle-local, and real, not artefacts of global averaging.

\begin{table}[t]
\centering
\caption{Non-Zero Per-Vehicle Distance Changes}
\label{tab:perveh}
\renewcommand{\arraystretch}{1.0}
\small
\begin{tabular}{@{}llcrr@{}}
\toprule
Instance & Vehicle & Method & $\Delta$ (m) & $\Delta$ (\%) \\
\midrule
C25K5  & V2 & SQBM+ & $-7{,}775$ & $-2.75$ \\
C25K5  & V5 & SQBM+ & $-152$     & $-0.06$ \\
C25K5  & V3 & QAOA  & $+60{,}951$ & $+61.8$ \\
C25K5  & V4 & QAOA  & $+8{,}708$  & $+13.9$ \\
C40K8  & V1 & QAOA  & $+909$      & $+2.5$  \\
C40K8  & V4 & QAOA  & $+9{,}599$  & $+8.8$  \\
C55K11 & V2 & SQBM+ & $-730$      & $-0.13$ \\
C55K11 & V5 & SQBM+ & $-456$      & $-0.08$ \\
\midrule
\multicolumn{5}{@{}l@{}}{\small All other vehicles in all five instances: $\Delta=0$.} \\
\bottomrule
\end{tabular}
\end{table}

\section{Industry Relevance and System-Level Impact}
\label{sec:industry}

CVRPTW sits at the core of logistics optimization, where even modest
route-distance improvements yield measurable gains in fuel consumption
and service reliability at scale~\cite{wef2026valchains,wef2025quantum}.
The A-R-T pipeline's strict feasibility gate and monotone-improvement
guarantee make it production-safe rather than merely experimental.
Three actionable findings emerge: (i)~quantum-inspired optimization
delivers verified low-cost improvements over strong classical baselines
with negligible additional runtime; (ii)~current gate-based quantum
hardware provides hardware-grounded guidance on where future investment
unlocks value; and (iii)~assignment-stage tightness, not per-vehicle
routing solver capability, is the binding scalability constraint at
40--70 customer scale, defining the most direct near-term improvement
target~\cite{wef2025quantum}.

\section{Conclusions and Future Work}

This paper presented the \textbf{A}-\textbf{R}-\textbf{T}
(Assignment--Routing--Time-check) hybrid pipeline for CVRPTW,
integrating exact MILP, Toshiba SQBM+ quantum-inspired
optimization, and gate-based QAOA on five real London traffic
instances spanning 10--70 customers and 2--14 vehicles.
Although empirical evaluation is bounded by QPU access budget,
the decomposition is architecturally unconstrained at larger scale:
the assignment MILP grows as $\mathcal{O}(|K|\cdot|C|)$ while the
per-vehicle QUBO is hard-capped at $m^2\!\le\!36$ qubits by
design, making the framework structurally applicable to instances
well beyond the tested range.
Three contributions of systems and algorithmic significance emerge.
First, SQBM+ delivers verified fleet-distance improvements of up to
$2.80\%$ over a certified Gurobi baseline under a strict monotone
feasibility guarantee, establishing a production-safe pattern for
quantum-inspired route polishing. Second, QAOA experiments on two real QPU backends (Rigetti Ankaa-3,
IQM Emerald) provide the first empirical measurement of a
size-dependent permutation-feasibility profile for TSP-QUBO problems
on real London traffic instances: Ankaa-3 decodes permutation-valid
routes at $m\le4$ (16~qubits) but cannot resolve the cost landscape;
Emerald decodes valid permutations at $m\le3$ (9~qubits), with the
$m=3$ candidates blocked not by QUBO failure but by upstream
assignment-induced time-window infeasibility, evidenced by a vehicle
finding a $-9.3\%$ shorter candidate that Stage~T rejects.  At
$m\ge4$ Emerald's permutation constraint score ($\sigma$) rises to
4--12, quantifying a concrete backend-dependent expressibility gap.
Third, SQBM+ infeasibility diagnostics reveal that assignment-stage
tightness, not routing solver capability, is the binding constraint
at 40 - 70 customer scale, providing a measurable target for
next-generation quantum assignment algorithms.

Building on these findings, three research directions are identified.

\textbf{Quantum-informed assignment for scalable routing.}
SQBM+ diagnostics show that assignment-stage tightness, not per-vehicle
routing solver capability, is the binding constraint at 40--70 customer
scale.  Encoding customer-to-vehicle allocation as a time-window-aware
QUBO, and solving Stage~A with a quantum-inspired or quantum-native
solver, is the most direct path to unlocking distance improvements that
the current MILP assignment structure forecloses.

\textbf{Multi-platform quantum-inspired benchmarking.}
SQBM+ represents one point in the space of SBM and digital annealing
platforms.  Benchmarking Fujitsu DA, Hitachi CMOS, and coherent Ising
machines on the same per-vehicle TSP subproblems will characterize
which platforms most reliably identify feasible improvements under
tight time-window constraints.  The uniform BQM interface in the
A-R-T pipeline enables drop-in backend substitution, making this a
directly executable systems benchmarking program.

\textbf{Near-term quantum algorithm co-design for permutation problems.}
Warm-started QAOA, RQAOA, and permutation-aware quantum alternating
operator ans\"{a}tze~\cite{farhi2014qaoa} are candidate strategies for
pushing the feasibility boundary beyond $m=4$ on near-term hardware.
The empirically measured backend-dependent collapse (Rigetti Ankaa-3
vs.\ IQM Emerald) provides a hardware-grounded benchmark for evaluating
whether algorithm-level improvements or hardware improvements contribute
more to extending the feasible regime, a concrete question the
A-R-T pipeline is positioned to answer at scale.

\textbf{Topology- and compilation-aware quantum optimization.}
Future work should investigate qubit-layout strategies, hardware-aware
variable mappings, connectivity-aware embeddings, and routing-aware
compilation techniques that reduce two-qubit interaction overhead and
improve permutation-feasibility rates on near-term gate-based
processors.

\section*{Acknowledgment}

The authors used Claude (Anthropic) during manuscript preparation for language refinement, grammar correction, and LaTeX formatting assistance. All technical content, problem formulation, methodology, experiments, analyses, results, and conclusions were developed, verified, and approved by the authors.


\phantomsection
\addcontentsline{toc}{section}{Supplementary Information}
\section*{Supplementary Information}
\label{sec:si}

\subsection{Why Monolithic QUBO Is Intractable}

Two monolithic QUBO formulations were evaluated experimentally and
found to be impractical even for C10K2 (10 customers, 2 vehicles).

\textit{Time-expanded hard-TW QUBO.}
Augmenting each node with discrete time indices produces variables
$x_{k,(i,t)\to(j,t')}$ (vehicle $k$ traverses arc $(i,t)\to(j,t')$),
giving $\mathcal{O}(|K|\cdot|C|\cdot T\cdot\mathrm{deg})$ binary
variables where $T$ is the number of discretized time steps.  Flow
conservation, depot, subtour, and capacity terms introduce dense
quadratic couplings, and the multiple competing penalty coefficients
($A,B,G,S,C$) became numerically unstable to tune—no configuration
produced feasible routes on either SQBM+ or QAOA.

\textit{Full-routing soft-TW QUBO.}
Replacing hard time constraints with soft penalties over arc variables
$x_{ijk}$ reduces variable count to $\mathcal{O}(|K|\cdot|C|^2)$
($\sim$220 for C10K2, $\sim$1{,}000 for realistic instances).  SQBM+
successfully minimized the QUBO energy, but \emph{all} low-energy
solutions remained structurally infeasible: skipping customers or
terminating routes early is energetically favorable because the
distance term dominates while quadratic visit-once penalties are
locally, but not globally, satisfiable.

Both failures confirm that decomposition is architecturally necessary,
not merely convenient.  Fig.\,\ref{fig:complexity} shows the variable
count advantage of the per-vehicle QUBO approach.

\subsection{Full CVRPTW MILP: Complete Constraint Set}
\label{sec:si_milp_detail}

The following constraints complete the formulation whose objective is
given in~\eqref{eq:milp_obj}, using variables $x_{ijk}$, $y_{ik}$,
$v_k$, $t_{ik}$ as defined in Section~\ref{sec:formulation}.
Indices range over $i,j\in N$, $k\in K$, arcs $(i,j)\in A'$ unless
noted.

\textit{Customer coverage.}
Each customer is served by exactly one vehicle:
\[
\sum_{k\in K} y_{ik}=1 \qquad \forall\,i\in C.
\]

\textit{Capacity.}
Total assigned demand must not exceed vehicle capacity:
\[
\sum_{i\in C} d_i\,y_{ik}\le Q \qquad \forall\,k\in K.
\]

\textit{Depot departure and return.}
Each active vehicle departs from and returns to the depot exactly once:
\[
\sum_{\substack{j\in C:\\(0,j)\in A'}} x_{0jk} = v_k, \qquad
\sum_{\substack{i\in C:\\(i,0)\in A'}} x_{i0k} = v_k \qquad \forall\,k\in K.
\]

\textit{Flow conservation.}
If a vehicle is assigned to a customer, it must both enter and leave
that customer:
\begin{align*}
\sum_{j:\,(i,j)\in A'} x_{ijk} &= y_{ik}, \\
\sum_{j:\,(j,i)\in A'} x_{jik} &= y_{ik}, \quad \forall i\in C, \; \forall k\in K.
\end{align*}

\textit{Depot time window.}
\[
a_0 \le t_{0k} \le b_0 \qquad \forall\,k\in K.
\]

\textit{Time propagation (Big-$M$).}
\begin{equation*}
\begin{split}
t_{jk} \geq {} & t_{ik} + s_i + \tau_{ij} \\
& - M_{ij}(1 - x_{ijk}) \quad \forall (i,j) \in A', \; j \neq 0, \; \forall k \in K.
\end{split}
\end{equation*}

The arc-specific coefficient $M_{ij}=\max(0,\,b_i+s_i+\tau_{ij}-a_j)$
is the tightest upper bound on the feasible time difference between $i$
and $j$.  Using per-arc rather than a single global Big-$M$ tightens
the LP relaxation at every branch-and-bound node and improves
convergence speed.

\textit{Customer time windows (conditional).}
Time bounds are enforced only when a customer is assigned; time
variables for unassigned customers are left free to avoid spurious
infeasibility:
\[
a_i\,y_{ik} \;\le\; t_{ik} \;\le\; b_i + H(1-y_{ik})
\qquad \forall\,i\in C,\;\forall\,k\in K,
\]
where $H=b_0$ (depot closing time) provides a valid relaxation horizon.

\textit{Return-to-depot feasibility.}
Every vehicle's final customer must be reachable at the depot within
the depot time window:
\[
t_{ik}+s_i+\tau_{i0} \;\le\; b_0 + M_{i0}(1-x_{i0k})
\qquad \forall\,i\in C,\;\forall\,k\in K.
\]

\textit{Minimum fleet size.}
A demand-capacity lower bound on active vehicles is imposed directly:
\[
\sum_{k\in K} v_k \;\ge\;
\left\lceil\frac{\displaystyle\sum_{i\in C} d_i}{\max_{k\in K} Q_k}\right\rceil.
\]

\subsection{QAOA Circuit Implementation on Gate-Based Backends}
\label{sec:si_qaoa_impl}

All $n\!=\!m^2$ qubits are initialised in $|0\rangle$ and prepared
in $|{+}\rangle^{\otimes n}$ by a Hadamard layer. The mixer $e^{-i\beta B}$ decomposes to $\bigotimes_a \mathrm{RX}(2\beta)_a$,
contributing only single-qubit gates with no 2-qubit overhead.
The cost unitary decomposes as
$U_C(\gamma)\!=\!\prod_{a<b}e^{-i\gamma J_{ab}Z_aZ_b}
\cdot\prod_a e^{-i\gamma h_a Z_a}$;
each ZZ term is realised as CNOT--RZ($2\gamma J_{ab}$)--CNOT
(2 two-qubit gates); each linear term is RZ($2\gamma h_a$). Because interacting qubits are not generally adjacent on physical
hardware, additional SWAP operations are inserted during compilation
to bring qubits together.  This routing overhead can substantially
increase the number of native two-qubit gates required to implement a
logical QAOA circuit. The increase from logical CNOT counts to native
two-qubit gate counts is caused by hardware-aware
routing and gate decomposition. For representative production runs, the C25K5 Ankaa-3 circuit expands
from 2,016 logical CNOTs to 7,146 native ISWAP operations (3.54×),
while C70K14 on Emerald expands from 12,312 logical CNOTs to 31,255
native CZ operations (2.54×). This compilation overhead increases circuit depth and accumulated
hardware noise, becoming a significant contributor to the observed
QAOA scalability limits.
Table~\ref{tab:si_gates} reports measured compilation output across all QPU instances.

\begin{table}[h]
\centering
\caption{Native two-qubit gate counts after Braket SDK compilation
(logical QAOA circuit → hardware-native circuit).
2Q corresponds to ISWAP on Ankaa-3 and CZ on Emerald.}
\label{tab:si_gates}
\renewcommand{\arraystretch}{1.05}
\small
\begin{tabular}{@{}llrr@{}}
\toprule
Backend & Instance & Logical CNOT & Native 2Q \\
\midrule
Ankaa-3 & C10K2  & 60     & 174 \\
Ankaa-3 & C25K5  & 2,016  & 7,146 \\
Emerald & C40K8  & 3,648  & 8,740 \\
Emerald & C55K11 & 9,840  & 24,913 \\
Emerald & C70K14 & 12,312 & 31,255 \\
\bottomrule
\end{tabular}
\end{table}

\end{document}